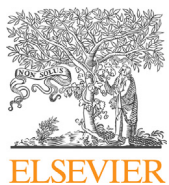
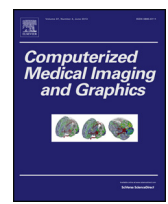

# Automated coronary artery atherosclerosis detection and weakly supervised localization on coronary CT angiography with a deep 3-dimensional convolutional neural network

Sema Candemir*, Richard D. White, Mutlu Demirer, Vikash Gupta, Matthew T. Bigelow, Luciano M. Prevedello, Barbaros S. Erdal

*Laboratory for Augmented Intelligence in Imaging of the Department of Radiology, The Ohio State University College of Medicine, United States*

## ABSTRACT

We propose a fully automated algorithm based on a deep learning framework enabling screening of a coronary computed tomography angiography (CCTA) examination for confident detection of the presence or absence of coronary artery atherosclerosis. The system starts with extracting the coronary arteries and their branches from CCTA datasets and representing them with multi-planar reformatted volumes; pre-processing and augmentation techniques are then applied to increase the robustness and generalization ability of the system. A 3-dimensional convolutional neural network (3D-CNN) is utilized to model pathological changes (e.g., atherosclerotic plaques) in coronary vessels. The system learns the discriminatory features between vessels with and without atherosclerosis. The discriminative features at the final convolutional layer are visualized with a saliency map approach to provide visual clues related to atherosclerosis likelihood and location. We have evaluated the system on a reference dataset representing 247 patients with atherosclerosis and 246 patients free of atherosclerosis. With five fold cross-validation, an Accuracy = 90.9%, Positive Predictive Value = 58.8%, Sensitivity = 68.9%, Specificity of 93.6%, and Negative Predictive Value (NPV) = 96.1% are achieved at the artery/branch level with threshold 0.5. The average area under the receiver operating characteristic curve is 0.91. The system indicates a high NPV, which may be potentially useful for assisting interpreting physicians in excluding coronary atherosclerosis in patients with acute chest pain.

## 1. Introduction

Coronary artery disease (CAD) represents the accumulation of atherosclerotic plaque within the walls of the coronary artery "tree". While this may cause restricted blood flow to the heart muscle (aka myocardium) from significant degrees of narrowing (aka stenosis) of the coronary artery lumen (Conti et al., 1978), even mildly stenotic plaque presents significant risk to the affected patient (Chang et al., 2018). Coronary computed tomography angiography (CCTA) is a well-established non-invasive imaging modality for evaluating chest-pain patients at low-intermediate risk for CAD (Rybicki et al., 2016). However, interpreting CCTA examinations for the presence or absence of coronary artery atherosclerosis is a time-consuming labor-intensive process and requires interpretive expertise. An Artificial Intelligence (AI)-based system that automatically extracts and helps to analyze coronary arteries and their branches, as demonstrated on CCTA, could be a useful clinical decision-making support tool to physicians interpreting these examinations (Hong et al., 2019).

Several successful automated and semi-automated algorithms to extract the artery centerlines, analyze the vessels, and detect the visual clues related to pathological changes related to atherosclerosis have been previously described. Some of these algorithms were developed explicitly to detect specifically calcified plaques (Wolterink et al., 2015) or plagues in general (Denzinger et al., 2019). Although these algorithms report high scores for abnormality detection, they are generally rule-based or based on conventional machine learning, such as by extracting shape and intensity-based features and then classifying the candidate regions using k-nearest neighbor or support vector machine classifiers (Wolterink et al., 2015).

Currently, the performance of deep learning-based methods surpasses that of conventional machine learning approaches for several computer vision problems (Krizhevsky et al., 2012); this

* Corresponding author.
*E-mail address:* Sema.Candemir@osumc.edu (S. Candemir).

2S. Candemir, R.D. White, M. Demirer et al. / Computerized Medical Imaging and Graphics 83 (2020) 101721

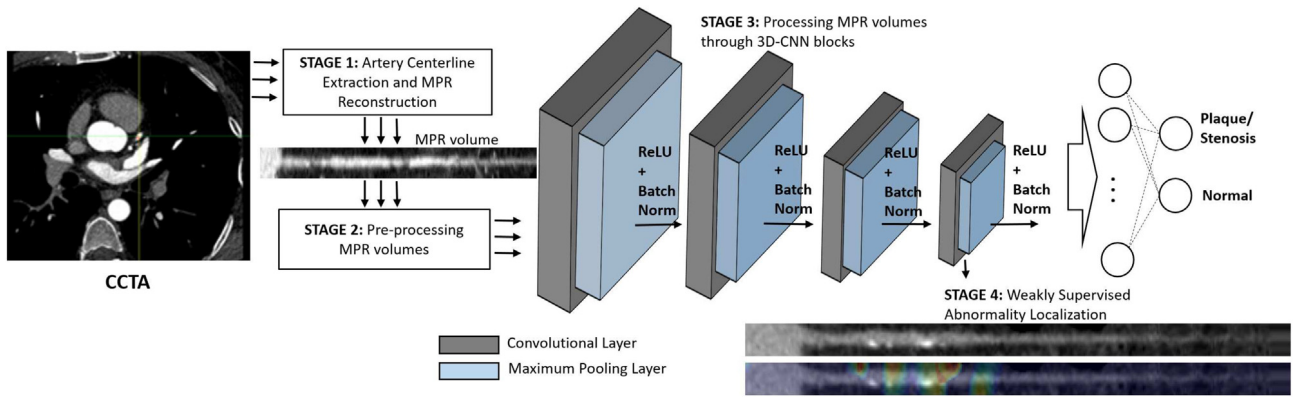

**Fig. 1.** Illustrative flowchart of the stages of the system pipeline. Stage 1: Extracting the major coronary arteries and branches on CCTA and representing each of them as a MPR volume. Stage 2: Applying pre-processing techniques to increase the CNN performance for modeling the discriminative features. Stage 3: Processing MPR volumes throughout the layers of the 3D-CNN. The system labels vessel volumes as normal and abnormal. Stage 4: Localizing the discriminative regions with a weakly supervised localization approach (CCTA: coronary computed tomography angiography, CNN: convolutional neural network, MPR: multi-planar reformation, ReLU: rectified linear units).

high performance of deep learning-based systems depends on the availability of large amounts of annotated data for training the model. Our aim in this study is to propose a deep-learning-based framework for screening CCTA examinations for the presence versus absence of coronary atherosclerosis. To that end, we utilize a 3-dimensional convolutional neural network (3D-CNN) to perform the image analysis. The system is trained with vessel volumes extracted from CCTA examinations of patients with or without atherosclerosis per coronary artery/branch. Thus, the network layers learn the discriminative visual characteristics of vessels hierarchically from a lower level to more abstract level. The trained system identifies the presence of atherosclerosis in coronary arteries and weakly localizes the pathological changes (e.g., plaques with and without vascular calcifications) based on the likelihood of the disease. Unlike prior studies (Wolterink et al., 2015; Denzinger et al., 2019), our approach is not designed for precise abnormality localization. Rather, the trained model highlights the saliency regions that contributing to the neural network algorithm's decision without any additional computation. The highlighted regions help to interpret and rationalize the decision of the model and, when the system is used in clinical practice, get interpreting physician's attention.

Usually, for a given model, the performance of a deep-learning-based system depends on the availability of large amounts of annotated training data. With efficient data curation (Erdal et al., 2018) and annotation (Demirer et al., 2019), our deep learning-based system can be integrated seamlessly into a clinical imaging workflow, allowing the training data to be updated regularly with new examples of CCTA examinations. Therefore, unlike the rule-based or conventional machine learning algorithms, our deep learning-based system is expected to prospectively process additional data, which helps the algorithm improving gradually for future clinical use.

Our contributions: (1) a deep-learning-based system is proposed that classifies coronary arteries as normal or abnormal, with visual display of likelihood of atherosclerosis presence per coronary artery/branch; (2) the system is fully automated, requiring no hand-crafted features or manual intervention; and (3) the system performance is initiated by applying random weights from scratch without employing any pre-trained model via transfer learning. To our knowledge, this is one of the earliest studies which utilizes a 3D-CNN architecture and shows the learned behavior of architecture towards classifying coronary arteries using visualization algorithms.

We have evaluated the system on our reference dataset (c.f., Section III.A), which contains 247 patients with atherosclerosis, and 246 patients with no atherosclerosis. With five fold cross-validation, an accuracy of 90.9%, with a Positive Predictive Value (PPV) of 58.8%, a Sensitivity of 68.9%, but a strong Specificity of 93.6%, and even stronger Negative Predictive Value (NPV) of 96.1% are achieved at artery/branch level, with decision threshold set at 0.5. The average Area Under the receiver operator characteristic Curve (AUC) is 0.91. Due to high NPV, the system may be potentially useful for assisting interpreting physicians in excluding atherosclerosis and identifying patients that need no additional medical testing or hospitalization (Litt et al., 2012).

The remainder of our report is organized as follows: Section 2 describes the proposed system containing (i) extraction of coronary arteries, (ii) pre-processing multi-planar reformation (MPR) volumes, (iii) processing volumes through a 3D-CNN architecture, and (iv) weakly supervised abnormality localization, with visual clues to help interpretation regarding atherosclerosis presence; Section 3 describes the data and reference standards, reports the implementation details, provides evaluation scores, and discusses the experimental results; Section 4 provides a conclusion.

## 2. Methodology

We propose a 3D CNN based supervised system which: (1) processes coronary artery volumes extracted from CCTA examinations; (2) characterizes the pathological structures and lesion; and (3) automatically locates the regions for providing visual clues for atherosclerosis. The proposed system is illustrated in Fig. 1 and has four stages: (i) extracting coronary arteries and branches from CCTA image datasets; (ii) pre-processing coronary artery volumes; (iii) classifying the vessels with the 3D CNN; and (iv) localizing the discriminative region for abnormal cases.

### 2.1. Stage 1: Extracting coronary arteries from CCTA

The initial stage of the system involves locating the coronary arteries and their branches within the CCTA image datasets. The coronary arteries/branches are represented by MPR volumes, which are reconstructed along the centerline of the vessels. Researchers proposed several successful algorithms for vessel detection (Gao et al., 2019a), vessel segmentation (Gao et al., 2019b,a), and vessel-centerline extraction (Gülsün et al., 2016; Wolterink et al., 2019). A comprehensive review of coronary artery detection approaches can be found in Lesage et al. (2009). We utilized the centerline extraction method in Zheng et al. (2013) due to its availability and its success on our CCTA sequences. The algorithm extracts the centerlines by deforming a mean shape model to the vessel volume with the coronary ostia and cardiac chambers as anchor points. This initial estimate of the coronary path is then refined through dynamic programming constrained with vessel-specific region-of-interest masks. We define the vessels as volumes around the centerline, as well as the surrounding area of the vessel in order to model the arterial characteristics, as well as those of surrounding anatomy. Ultimately, straightened MPR volumes in a longitudinal view of vessels are produced (Fig. 2).

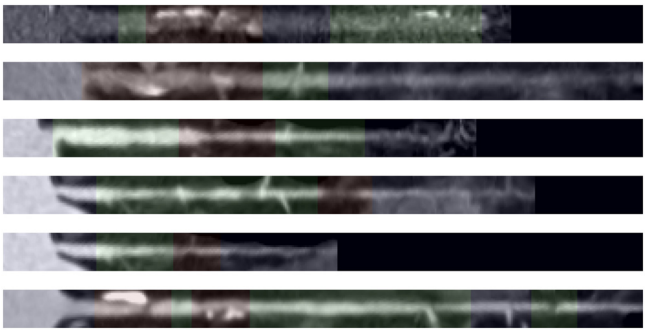

**Fig. 2.** Straightened MPR volumes in longitudinal view of a coronary vessel branch are overlapped by severity labels (severe = red; mild = green) (for ease of visualization, only central frames of each volume are shown). (For interpretation of the references to color in this figure legend, the reader is referred to the web version of this article.)

### 2.2. Stage 2: Pre-processing MPR volumes for Training

After the extraction of the coronary arteries/branches and building MPR volumes, we apply pre-processing techniques to the arterial volumes to increase the system's classification and localization performance. We first utilize a clamping technique that restricts the processing intensity into a predetermined range. With $I(x, y)$ representing CT intensity value at position $(x, y)$, we clamp intensities in the MPR volume as:

$$I(x, y) = \begin{cases} 800, & \text{if } I(x, y) < 800 \\ I(x, y), & \text{if } 800 < I(x, y) < 2200 \\ 2200, & \text{if } I(x, y) > 2200 \end{cases} \quad (1)$$

where $I(x, y)$ is in the range [0 4095] CT values in a 12-bit CCTA scanner. After clamping, we normalize the volumes over the dataset to help the system with faster and robust convergence.

The extracted coronary arteries/branches have different lengths; however, the MPR volumes need to be resized for sequential processing through CNNs. The shape of the vessel borders is one of the significant visual clues for stenosis; therefore, we do not apply the interpolation technique for resizing. Instead, we extend the volumes by adding empty frames to the end of the arterial volume. The processing length of the arterial volumes is decided based on the longest annotated vessel in the dataset.

The voxel-based CNNs are prone to overfitting due to high dimensional data combined with lack of annotated subjects to optimally train a large number of parameters (Litjens et al., 2017). To mitigate potential overfitting, we accumulate the training data by randomly rotating the MPR volumes between 0° and 360° around the vessel centerlines. The number of extracted arteries/branches is imbalanced across the classes, therefore, we apply different levels of augmentation to balance the classes in the training set.

### 2.3. Stage 3: Model configuration of 3D convolutional neural network

Each coronary MPR volume in the dataset is processed through a 3D-CNN architecture by convolving the volume with a set of kernels. The network architecture consists of four convolutional layers with kernels of $3 \times 3 \times 3$ elements, with 32, 64, 128, and 256 filters. The number of convolutional layers is decided empirically based on the performance of the model on the validation set. A $2 \times 2 \times 2$ max-pooling layer follows each convolutional layer for parameter reduction and spatial invariance. The network layers capture the spatial imaging characteristics of MPR volumes through their inherent mechanism hierarchically from low level to more abstract features and learn the discriminative features between the vessel fragments with and without atherosclerosis. The architecture uses the rectified linear units (ReLU) activation function, which introduces non-linearity to the system (LeCun et al., 2015). Each activation function is followed by a batch normalization, which mitigates the overfitting and improves the system generalization by normalizing the output of the activation function (Ioffe and Szegedy, 2015). The output of the deepest convolutional layer is flattened and fed to the fully connected layer, which serves as a classifier in the architecture and processes features extracted through convolutional layers. The fully connected layer continues with a dropout layer (Srivastava et al., 2014) in which the system temporarily ignores the randomly selected neurons during the training to prevent the system from memorizing the training data. We have also used $L2$ weight decay to penalize the model to increase the generalization capacity of the model. The parameter values and the training strategy of the 3D-CNN are reported in Section 3.2.

### 2.4. State 4: Weakly supervised abnormality localization

Several methods have been proposed for visualizing and understanding the learned behavior of CNNs (Fong and Vedaldi, 2017; Simonyan et al., 2013; Zhou et al., 2016; Selvaraju et al., 2017). The saliency approaches, including class activation map (CAM) (Zhou et al., 2016) and gradient-based class activation map (Grad-CAM) (Selvaraju et al., 2017), identify the discriminative regions by constituting a weighted-feature layer ("heatmap") highlighting the discriminative features learned by the CNN.

In this study, we utilize Grad-CAM (Selvaraju et al., 2017) in which the gradient information of the image flowing into the last convolutional layer that produces a map highlighting the importance of each pixel on the image. Our aim in deriving saliency maps is to interpret and rationalize the decision of the trained system. If a model correctly predicts the coronary artery/branch as diseased, we expect the system to highlight (localize) the pathological changes (abnormalities, such as atherosclerosis plaques). The highlighted clues to the presence of an abnormality confirm that the model made its decision based on the discriminative features (pathological abnormalities), not on some arbitrary parameter. Therefore, highlighting serves as visible indicators as to what the system has learned and uses to assign the vessel volume to one of the classes. For diseased cases, we would expect the system to highlight the locations for the presence of atherosclerosis.

## 3. Experiments and results

### 3.1. Dataset

Under prior approval from the Institutional Review Board with the waiver of informed consent, we retrospectively collected CCTA examinations of 493 patients acquired at the Ohio State University Wexner Medical Center between March 2013 and July 2018. The examinations were performed using multi-detector CT systems (Siemens Healthineers). During the CCTA image curation process for algorithm development, the patient's personal and health information is de-identified (Health Insurance Portability and Accountability Act). In the curated dataset, 247 patients demonstrate atherosclerosis, and 246 patients are atherosclerosis-free based on review by our investigator-expert (RDW with 33-year experience in cardiovascular imaging and ACC/AHA Level III CCT certification).

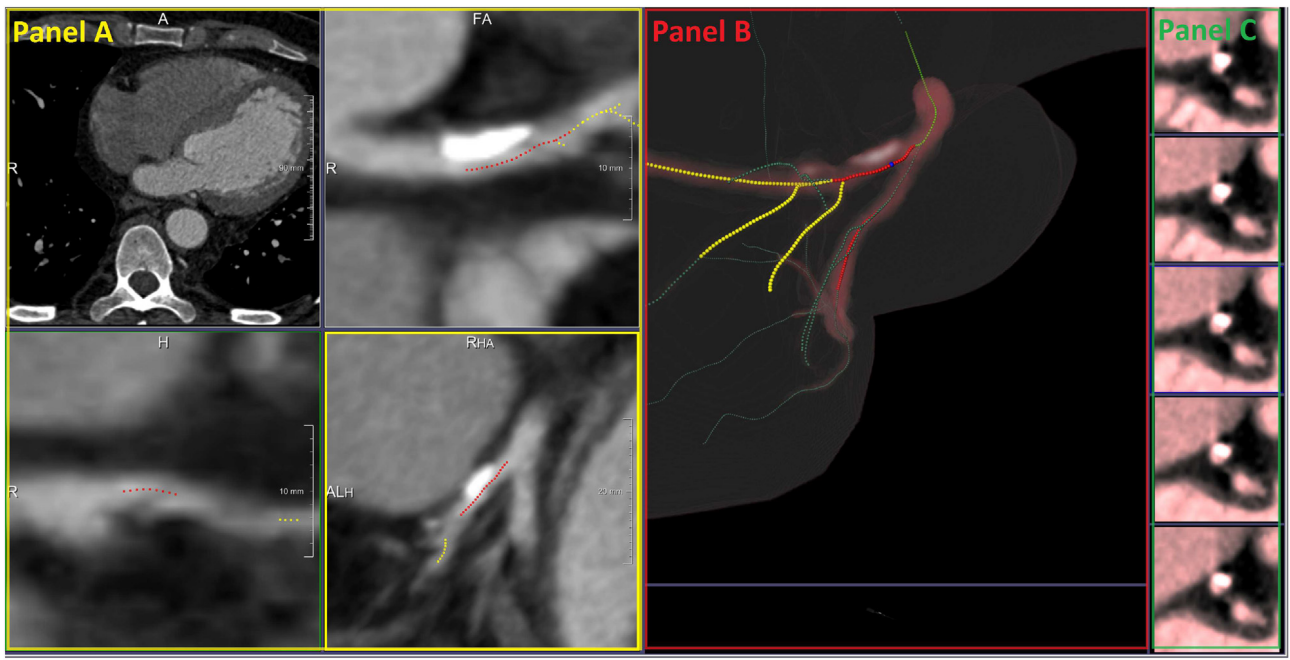

**Fig. 3.** Screenshot of the graphical user interface for coronary artery vessel detection and expert image annotation. Panel A. Axial (upper left) and multi-planar reconstructions. The proximal segment of the left anterior descending coronary artery is manually annotated to indicate severe (red) due to the presence of prominent atherosclerotic plaque causing flow-limiting (>50%) luminal narrowing. More distal segments and diagonal branches demonstrating milder degrees of plaque and narrowing are correspondingly labeled (yellow). Panel B. Volume-rendered cardiac display along with coronary artery map. The interface allows the expert to actively engage in the annotation process, with an indicator of the active level (blue marker), as well as the atherosclerosis severity level as severe (red) or mild (yellow); segments without atherosclerosis are also indicated (green). Panel C. Sequential short-axis (intravascular ultrasound (IVUS)-like) views of the target coronary artery/branch. The central sub-panel corresponds to the active level (i.e., blue marker), while top sub-panels are proximal and bottom sub-panels distal in the same artery/branch. The expert examines and validates the plaque characteristics, including associated luminal narrowing using short-axis views. (For interpretation of the references to color in this figure legend, the reader is referred to the web version of this article.)

One of the challenges in developing a deep-learning model for medical-image analysis is curating high-quality annotated data as a reference standard to train and validate the system. The standard clinical tools used for coronary artery evaluation do not have annotation capability; therefore, we have designed an annotation interface (Demirer et al., 2019), which has visualization and manipulation capabilities that are similar to a clinical image interpretation system (syngo-via). The previously described interface (Demirer et al., 2019) is built on top of MevisLab (Ritter et al., 2011), which is an image processing and visualization medical-image analysis research and development platform.

The presence of atherosclerotic plaque and severity of the stenosis were manually marked by an expert cardiovascular imager (RDW) using the annotation interface (Demirer et al., 2019) by following the guidelines of the Society of Cardiovascular Computed Tomography (Cury et al., 2016). Diseased coronary segments fragments are graded as mild (i.e., <50% luminal narrowing) or severe (i.e., >50% luminal narrowing) (Fig. 3). Of the 493 CCTA examinations, 247 contained coronary artery atherosclerosis, while 246 of them were free of coronary artery atherosclerosis. Included in the subpopulation of 247 diseased cases, there were 641 coronary arteries or branches with atherosclerosis selected among the major coronary arteries: left anterior descending (LAD), left circumflex (LCx) and right coronary artery (RCA).

### 3.2. Implementation details and training strategy

The study dataset consists of 493 CCTA examinations, with 247 demonstrating atherosclerosis and 246 atherosclerosis-free. We performed five fold cross-validation to reduce performance differences due to the relatively small size dataset and to provide more robust generalization performance. At each fold, 60% of the dataset is used to train the model, 20% for model selection by validation, and 20% to test the model. The cross-validation is applied at the subject level to prevent the data leakage between train, test, and validation sets.

The CCTA image datasets are in Digital Imaging and Communications in Medicine (DICOM) format. We use simpleITK library (simpleITK) to convert DICOM volumes into NumPy arrays. The system is compiled with Adam optimizer (Kingma et al., 2014). The training parameters are listed in Table 1. As a training strategy, we monitor the model performance and use two early stopping callbacks to stop the training before the model begins to overfit (Goodfellow et al., 2016). We set a large epoch value 1000 as upper bound iteration. The number of training iteration is decided automatically based on the model performance on the validation and training set. If the validation loss has started to increase during the training process, the system triggers the early stopping callback. If the validation loss continues to increase for another 20 iterations, then the system stops the training. The continuous increase in validation loss is an indication of overfitting. The second callback is monitoring the training accuracy. If the training accuracy reaches to a maximum value (e.g., 0.99–1), the early stopping callback stops the training to prevent the model from memorizing the data (overfitting), since it indicates no further improvement. The weights are randomly initialized from scratch with random weights employing any pre-trained model via transfer learning.

**Table 1**
Implementation details - parameters.

| Parameter | Value |
| --- | --- |
| Processing dimension of each MPR volume | $21 \times 21 \times 350$ voxels |
| Optimizer | Adam (Kingma et al., 2014) |
| Learning rate | $10^{-6}$ |
| $\beta_1$ | 0.9 |
| $\beta_2$ | 0.999 |
| $\epsilon$ | $10^{-8}$ |
| Loss function | Categorical cross-entropy |
| Batch size | 32 |
| Drop-out keep rate | 0.5 |
| $L2$ weight decay (Krogh and Hertz, 1992) $\lambda$ | $10^{-3}$ |
| Early stopping max epoch | 1000 |
| Early stopping patience epoch | 20 |

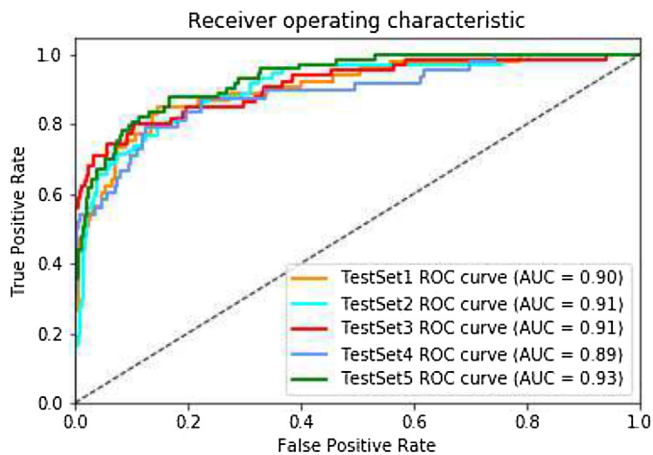

**Fig. 4.** Plot depicting system performance for vessel classification. The average area under the curve is 91% (AUC = area under the curve, ROC = receiver operator characteristic).

The model is developed in Python (version 3.6.8) using Tensorflow Keras API (version 2.1.6-tf) and trained on an Nvidia Quadro GV100 system with 32GB graphics cards and with CUDA/CuDNN v9 dependencies for GPU acceleration.

### 3.3. System classification performance evaluation

The classification performance of the system is evaluated at the coronary artery/branch level. We run the system at each fold, compute the system's prediction on test data, and average the evaluation metric scores across all the models. The performance metrics used in the study include Sensitivity, Specificity, Accuracy, PPV, NPV, and AUC.

Table 2 lists the performance metrics obtained with the proposed system in terms of mean and standard deviation across the cross-validated folds. The receiver operator characteristic (ROC) curves for each test fold is in (Fig. 4).. The system achieved an Accuracy of 90.9%, with a PPV of 58.8%, Sensitivity of 68.9%, Specificity of 93.6%, and NPV of 96.1% at threshold 0.5. The average AUC is 91%.

### 3.4. Comparison with the literature

There are successful algorithms in the literature proposed for vessel detection (Gao et al., 2019a), vessel segmentation (Gao et al., 2019b,a; Lesage et al., 2009), centerline extraction (Gülsün et al., 2016; Wolterink et al., 2019), and detecting and quantifying stenosis (Kirişli et al., 2013; Zreik et al., 2018). The comparison of the state-of-the-art algorithms proposed before 2013 can be found in (Kirişli et al., 2013). The novelty of our study is coronary artery classification (assigning binary labels as healthy or diseased per artery/branch) utilizing a 3D-CNN architecture and abnormality localization with saliency maps. Therefore, in this section, we only mention automated coronary artery classification and abnormality localization approaches. Table 3 provides a rough comparison of similar studies.

*Conventional machine learning algorithms:* In Mittal et al. (2010), a conventional machine learning approach was proposed that extracts features (e.g., intensity and gradient histograms) along the coronary artery centerline to train a probabilistic boosting tree and random forest algorithms with the extracted features. Similarly, Sankaran et al. (2016) proposed extraction of shape features and training of a random forest regression algorithm to predict the healthy lumen diameter. The main difference between our approach and conventional approaches for vessel classification is that our approach does not determine hand-crafted features. Instead, the system learns the most discriminative features between atherosclerosis and atherosclerosis-free classes during the training.

*Deep learning:* Currently, the performance of deep-learning-based methods surpass the performance of conventional machine learning approaches for several computer vision problems (LeCun et al., 2015). However, the performance of deep neural network architectures depends on the use of large annotated datasets. The well-annotated high-quality large training data is limited in the medical imaging domain since the data curation process is resource- and time-intensive. Moreover, expert knowledge is necessary for annotations (e.g., only an expert in cardiovascular imaging can confidently annotate the coronary artery tree for atherosclerotic changes and decide the reference standards). In such a scenario, transfer learning can boost performance by providing pre-trained architectures (e.g., VGG-16 Simonyan and Zisserman, 2014), which previously trained with 1.2 million general images with 1000 categories from ImageNet (Krizhevsky et al., 2012). The transfer learning strategy is easily applicable to 2-Dimensional (2D) images due to the abundance of suitable pre-trained models. However, this is not the case for 3D images, as is commonly found in advanced forms of imaging (e.g., CCTA), due to lack of pre-trained models applicable to 3D volumes. Therefore, in our previous study (Gupta et al., 2019), we obtained 2D projections of vessel volumes and modeled each vessel as a 2D frame, which contains different projected views of the coronary artery/branch. Although we successfully utilized the transfer learning strategy from 2D images to 3D by obtaining the projected views of the volumes, the projected views are sparse representation of volumes and contain less information than processing the whole vessel. Therefore, in this study, we analyze the vessel structure in 3D space that allows us considering all viewing angles of the vessel structure.

Another recent study, which analysis coronary arteries through a deep neural network system, utilizes a 3D CNN to extract image features for each 25 × 25 × 25 volume segments (cubes) along the MPR volume of the artery and processes the sequential features with a recurrent neural network (Zreik et al., 2018). The system is proposed to detect plaque and stenosis, as well as determining plaque-type and significance of a stenosis. The study contains limited training dataset (CCTA from 98 patients) compared to our study; therefore, a shallower 3D CNN is employed to analyze the segments, which have fewer parameters and are, therefore, less prone to overfitting. However, the recurrent neural network maintains the context of the sequential segments and leverages the success of the system for stenosis detection and grading.

More recently, another deep-learning-based algorithm has been proposed in Denzinger et al. (2019). The system modifies the approach presented in Zreik et al. (2018) by transforming the volume segments (cubes) into the polar coordinate system and extracting features frame-by-frame. In addition, they have extracted radiomics features of each volume segment and combine these features with deep features.

*Our study:* We investigate the vessel classification performance of 3D CNN by processing the whole vessel. To our knowledge,

**Table 2**

The average system performance. Coronary artery/branch-based statistical results from testing (FN: False Negative, FP: False Positive, NPV: Negative Predictive Value, PPV: Positive Predictive Value, TN: True Negative, TP: True Positive).

| | Formula | Metric th = 0.1 | Metric th = 0.2 | Metric th = 0.3 | Metric th = 0.4 | Metric th = 0.5 |
|---|---|---|---|---|---|---|
| Accuracy | (TP+TN)/(TP+FP+FN+TN) | 0.931 ± 0.017 | 0.924 ± 0.020 | 0.921 ± 0.021 | 0.916 ± 0.019 | 0.909 ± 0.018 |
| PPV | TP/(TP+FP) | 0.805 ± 0.173 | 0.712 ± 0.134 | 0.669 ± 0.113 | 0.624 ± 0.101 | 0.588 ± 0.080 |
| Sensitivity | TP/(TP+FN) | 0.567 ± 0.073 | 0.619 ± 0.073 | 0.658 ± 0.083 | 0.681 ± 0.095 | 0.689 ± 0.095 |
| Specificity | TN/(FP+TN) | 0.977 ± 0.023 | 0.963 ± 0.026 | 0.954 ± 0.028 | 0.944 ± 0.028 | 0.936 ± 0.027 |
| NPV | TN/(FN+TN) | 0.947 ± 0.010 | 0.953 ± 0.010 | 0.957 ± 0.009 | 0.960 ± 0.009 | 0.961 ± 0.008 |
| | Formula | Metric th = 0.6 | Metric th = 0.7 | Metric th = 0.8 | Metric th = 0.9 | |
| Accuracy | (TP+TN)/(TP+FP+FN+TN) | 0.902 ± 0.022 | 0.888 ± 0.022 | 0.875 ± 0.024 | 0.844 ± 0.030 | |
| PPV | TP/(TP+FP) | 0.557 ± 0.075 | 0.506 ± 0.066 | 0.469 ± 0.068 | 0.404 ± 0.043 | |
| Sensitivity | TP/(TP+FN) | 0.714 ± 0.096 | 0.724 ± 0.091 | 0.761 ± 0.091 | 0.796 ± 0.083 | |
| Specificity | TN/(FP+TN) | 0.924 ± 0.032 | 0.907 ± 0.032 | 0.888 ± 0.035 | 0.848 ± 0.041 | |
| NPV | TN/(FN+TN) | 0.964 ± 0.007 | 0.964 ± 0.007 | 0.968 ± 0.008 | 0.972 ± 0.007 | |

**Table 3**

ML: machine learning, 2D: 2-dimension, 3D: 3-dimension, CNN: convolutional neural network, R-CNN: recurrent convolutional neural network, Acc: accuracy, Sens: sensitivity, Spec: specificity, AUC: area under curve, CCTA: coronary computed tomography angiography, Grad-CAM: gradient-based class activation map.

| Study | Method | Architecture | Processing size | # of images | Abnormality detection |
|---|---|---|---|---|---|
| Mittal et al. (2010) | Conventional ML Feature Engineering Supervised ML | Intensity + Gradient boosting tree random forest | Artery Segments | 165 patients 355 calcified lesions | Evaluated on segment-level only calcification Acc: 81% |
| Sankaran et al. (2016) | Conventional ML Feature Engineering Supervised ML | Shape Features random forest regression | Artery Segments (called stems in the study) | 350 patients (10,697 sections) | Evaluated on segment-level Sens: 90%, Spec: 85% AUC: 0.90 |
| Gupta et al. (2019) | Deep Learning Transfer Learning | 2D-CNN Inception (Szegedy et al., 2015) | Artery Branch | 247 CCTA | Evaluated on artery-level AUC: 0.93 |
| Zreik et al. (2018) | Deep Learning Training from Scratch | 3D-R-CNN | Artery Segments (Cubes) | 163 patient | Evaluated on segment-level – Acc: 94%, F1: 80% artery-level – Acc: 93%, F1: 88% |
| Denzinger et al. (2019) | Deep Learning + Conventional ML | 3D- R-CNN Radiomics features | Artery Segments (Cubes) | 95 patient | Evaluated on segment-level AUC: 0.96, Acc: 92% |
| This study | Deep Learning Training from Scratch | 3D-CNN | Artery Branch | 247 CCTA | Evaluated on artery-level AUC: 0.91 Localization Grad-CAM (Selvaraju et al., 2017) Dice: 0.70%, Sens: 0.76% |

there is no available 3D pre-trained model suitable to apply the transfer learning strategy easily. Therefore, we perform training by initializing the weights randomly (training from scratch) instead of employing them from a pre-trained model; and train the system with a relatively small dataset. Experimental evaluation has demonstrated that training with small-sized datasets performs as good as fine-tuning a pre-trained model (see Table 3). Slightly lower performance in 3D CNN can be explained by the lack of data to find the optimal weight set. The voxel-based CNNs are prone to overfitting due to high-dimensional data in the setting of a relatively small number of annotated subjects to optimally train a large number of parameters (Litjens et al., 2017). Our aim is assigning binary labels as healthy or diseased per-artery/branch. Therefore, we use a simpler architecture compared to 3D-RCNN in Zreik et al. (2018) which is computationally more expensive than 3D CNN due to keeping track of the states in time. To our knowledge, we have the largest annotated dataset for patients demonstrating atherosclerosis (247 patients), that allows us to use slightly deeper architecture than the CNN architecture in Zreik et al. (2018). However, despite using the largest training dataset in the literature, applying a considerable amount of augmentation and regularization, our experiments indicated that the relatively smaller dataset is still limited in training a system from scratch and find the optimal weight set (note that the weights of pre-trained models are optimized through 1.2 million images (ImageNet Krizhevsky et al., 2012). On the other hand, our training dataset contains approximately 4000 positive and 4000 negative arteries/branches obtained by augmenting the available annotated vessels extracted from 247 subjects with CCTA examinations (c.f., Section 3.2).

### 3.5. Weakly supervised localization evaluation

The system localization assessment is performed on the artery/branch level. We apply the Grad-CAM technique (Selvaraju et al., 2017) (c.f., Section 2.4) and derived saliency maps ("heat maps" as visual clue to the interpreting physician. The higher "temperature" on the heat map (red and yellow-colored) represents the most discriminative location for the system to indicate atherosclerosis. We compared the system decision with the cardiovascular imager's annotations (c.f., Section 3.1). For a successful localization, we expect (i) the system to highlight the discriminative regions specific to atherosclerosis; and (ii) the system decision to be compatible with the expert imager's decision. We obtained heatmap for test cases, which are correctly classified as abnormal (218 vessel branches) (Fig. 5).

We first measure the overlapping at pixel level and compare the system decision with the reference annotation. For each pixel, if the system decision overlaps with expert annotation, it is True Positive (TP). If the expert imager annotated as positive but the system classified as negative, it is False Negative (FN). If the system decision is positive, but the expert did not annotate as diseased, it is False Positive (FP). Table 4 lists the overlapping scores at pixel level.

While the system heat map shows fairly discrete salient abnormal regions (e.g., the peak point of the atherosclerotic region), the expert imagers's annotation covers a larger area, starting at initialization of luminal narrowing till the end of luminal narrowing. Therefore, to be fair to the system decision on abnormal regions, we have applied regional evaluation. If the system saliency is within the imager's annotation, it is considered TP. If the system does not highlight an annotated area, it is FN. If there is a salient region in the wrong place (not in the annotated area), it is FP. Fig. 6 illustrates the mentioned cases. Table 5 lists the overlapping scores at the region-level. Out of 218 correctly classified test cases, the system successfully locates the regions demonstrating atherosclerosis (i.e., overlapping with physician annotation) on 191 arteries. However, the system also highlights areas resulting from artifacts and leading to FP determinations (105 arteries out of 218).

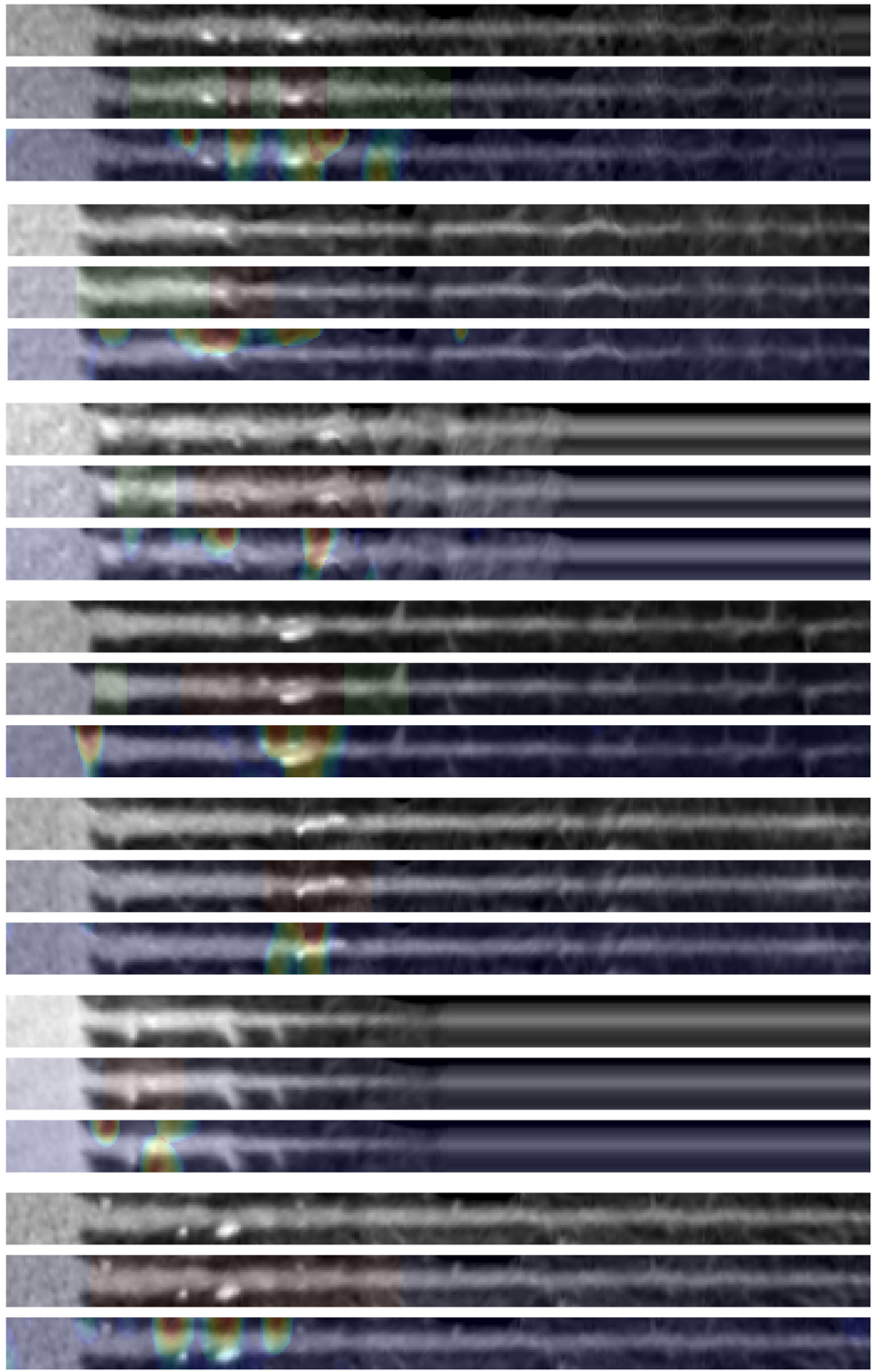

**Fig. 5.** Straightened MPR vessel with expert imager annotations for regions related to atherosclerosis, and system saliency maps. For each triad, the first frame is central frame of the vessel volume; the middle frame is the imager's annotation superimposed on the vessel (red = severe; green = mild); and the last frame shows the saliency map ("heat map") indicating the discriminative regions for the system and basis for assignment of class (atherosclerotic vs. atherosclerosis-free) (for ease of visualization, central frames of each volume are shown). (For interpretation of the references to color in this figure legend, the reader is referred to the web version of this article.)

**Table 4**
Weakly supervised localization evaluation: pixel-level overlap scores.

| Metric | Formula | Confusion matrix values | Calculated value |
|---|---|---|---|
| Dice | 2TP/2TP + FP + FN | 2 ∗ 19,629/2 ∗ 19,629 + 22,416 + 5778 | 0.58 |
| Accuracy | TP + TN/TP + TN + FN + FP | 19,629 + 28,477/19,629 + 28,477 + 5778 + 22,416 | 0.63 |
| Sensitivity | TP/TP + FN | 19,629/19,629 + 5778 | 0.77 |

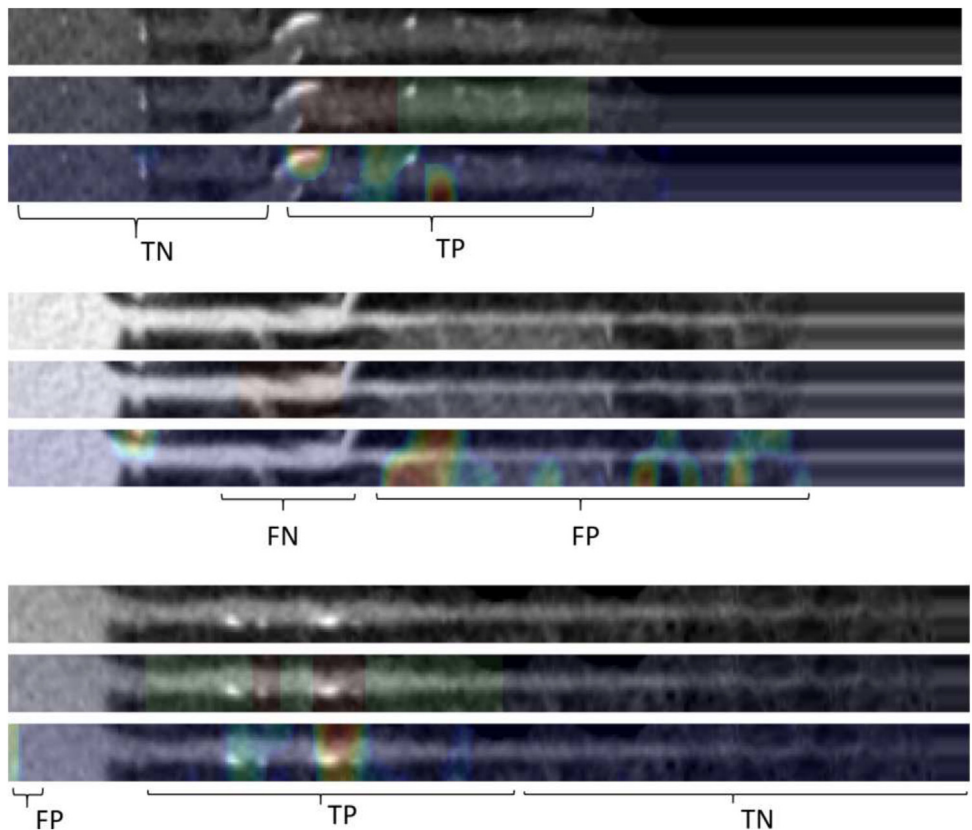

Fig. 6. Illustration of the saliency evaluation at region level.

Table 5
Weakly supervised localization evaluation: region-level overlap scores.

| Metric | Formula | Confusion matrix values | Calculated value |
| --- | --- | --- | --- |
| Dice | 2TP/2TP + FP + FN | 2 ∗ 191/2 ∗ 191 + 105 + 62 | 0.70 |
| Accuracy | TP + TN/TP + TN + FN + FP | 191 + 105/101 + 105 + 62 + 105 | 0.64 |
| Sensitivity | TP/TP + FN | 191/191 + 62 | 0.76 |

## 4. Discussion and conclusions

Interpreting CCTA examinations for clinical care is a time-consuming and labor-intensive process, and requires expertise in cardiovascular imaging. An AI-based system that automatically processes CCTA examinations, including the extraction of artery/branch centerlines, as well as detection and localization vs. exclusion of the atherosclerotic changes, could serve as a valuable tool assisting interpreting physicians (Rybicki et al., 2016; Hong et al., 2019; Litt et al., 2012). In this study, we propose a fully automated system that classifies coronary vessels as normal/abnormal and locates the discriminative regions related to any atherosclerosis. The entire process requires no hand-crafted features or human intervention. We utilized a 3D CNN to perform the image analysis of the coronary artery tree. The system performance is investigated by applying random-weights initialization without employing any pre-trained model. To our knowledge, this is one of the earliest studies which utilizes a 3D CNN architecture and shows the learned behavior of the architecture towards classifying the coronary vessels using visualization algorithms.

With five fold cross-validation, the average classification Accuracy of the system for coronary arteries is 90.9%, with limited strength for atherosclerosis detection (e.g., PPV = 58.8%), but strong ability to exclude atherosclerotic plaque (e.g., NPV = 96.1%), suggesting that it may potentially be useful for automatically clearing the patients clinically for avoidance of unwarranted additional diagnostic evaluations or hospital admissions (Litt et al., 2012). However, one of the potential reasons for the success in identifying non-diseased coronary arteries/branches might be the relatively higher number of negative vessels in the training data, requiring no augmentation, therefore, capturing more variations of negative cases.

Visualizing the behavior of a CNN (previously considered as a black-box machine learning technique) in providing interpretations as a system output is especially crucial in trusting medical diagnostic applications of AI (Zhang et al., 2017). Accordingly, the view of the US Food and Drug Administration (FDA) is that any clinical decision support software should provide a rationale for its decisions (FDA, 2017). Therefore, we visualize the learned behavior of our system in assigning a class label to coronary vessels with suspected atherosclerosis using saliency maps. We expect

(i) the system to highlight the discriminative regions specific to atherosclerosis; and (ii) the system decision to be consistent with the physician's decision. We compared the system decision with the expert imager's annotations and found generally strong agreement (Dice coefficient 0.70 at region level).

The encouraging performance of deep learning-based systems depends on the availability of large amounts of annotated data for training. In this study, we investigated the coronary artery/branch classification performance of a CNN-based system with training from-scratch strategy and with a relatively small-sized dataset. Based on our experimental results, the system performance is comparable with the performance obtained with pre-trained models with 2D images. In the Laboratory for Augmented Intelligence in Imaging, we built a seamless workflow to facilitate data curation and annotation (Erdal et al., 2018). We have also developed a clinician-friendly user interface to annotate the coronary arterial system (Demirer et al., 2019). With the integration of our model into the data curation workflow, the training data will be updated regularly with the new examples of CCTA examinations containing visual clues for atherosclerosis. Therefore, the system is expected to process additional data prospectively, which could potentially help to improve the algorithm for future clinical use.

## Authors' contribution

Sema Candemir: conceptualization, methodology, software, writing original draft, review and editing. Richard D. White: conceptualization, data curation, writing original draft, review and editing. Mutlu Demirer: conceptualization, software, data curation. Vikash Gupta and Luciano M. Prevedello: conceptualization, review and editing. Matthew T. Bigelow: conceptualization, data curation. Barbaros S. Erdal: conceptualization, data curation, review and editing.

## Acknowledgment

This research is supported by the Department of Radiology of The Ohio State University College of Medicine. In addition, the project is partially supported by donation from the Edward J. DeBartolo, Jr. Family (Funding), Master Research Agreement with Siemens Healthineers (Technical Support), and Master Research Agreement with NVIDIA Corporation (Technical Support). We thank the anonymous reviewers for their suggestions that have helped to improve the quality of the manuscript.

**Conflict of interest**: None declared.